\documentclass[fleqn,usenatbib]{mnras}

\usepackage{newtxtext,newtxmath}

\usepackage[T1]{fontenc}

\DeclareRobustCommand{\VAN}[3]{#2}
\let\VANthebibliography\thebibliography
\def\thebibliography{\DeclareRobustCommand{\VAN}[3]{##3}\VANthebibliography}

\usepackage{graphicx}	
\usepackage{amsmath}	
\usepackage{amssymb}	

\usepackage{epstopdf}
\usepackage{ulem}
\usepackage{amssymb}
\usepackage{times}
\usepackage{graphicx}
\usepackage[usenames,dvipsnames]{pstricks}
\usepackage{psfrag}
\usepackage{lipsum}
\usepackage{natbib}
\usepackage{mathtools} 

\def\araa{ARA\&A}
\def\apj{ApJ}
\def\apjs{ApJS}
\def\aap{A\&A}
\def\jcap{J. Cosmology Astropart. Phys.}
\def\mnras{MNRAS}
\def\prd{Phys.~Rev.~D}
\def\prl{Phys.~Rev.~Lett.}
\def\nat{Nature}

\newcommand{\be}{\begin{equation}}
\newcommand{\ee}{\end{equation}}
\newcommand{\bary}{\begin{eqnarray}}
\newcommand{\eary}{\end{eqnarray}}

\title[Electron-positron pair plasma in TXS 0506+056 and the ``neutrino flare" in 2014 - 2015]{Electron-positron pair plasma in TXS 0506+056 and the ``neutrino flare" in 2014 - 2015 }

\author[N. Fraija et al.]{
N. Fraija,$^{1}$\thanks{E-mail: nifraija@astro.unam.mx}
E. Aguilar-Ruiz,$^{1}$
A.~Galv\'an-G\'amez $^{1}$
\\
$^{1}$Instituto de Astronom\' ia, Universidad Nacional Aut\'onoma de M\'exico, Circuito Exterior, C.U., A. Postal 70-264, 04510 M\'exico City, M\'exico
}

\date{Accepted XXX. Received YYY; in original form ZZZ}

\pubyear{2015}

\begin{document}
\label{firstpage}
\pagerange{\pageref{firstpage}--\pageref{lastpage}}
\maketitle

\begin{abstract}
The detection of a prolonged flaring activity from blazar TXS 0506+056 in temporal and spatial coincidence with the energetic neutrino IceCube-170922A provided evidence about the photo-hadronic interactions in this source. However, analysis of the archival neutrino and multi-wavelength data from the direction of this blazar between September 2014 and March 2015 revealed a ``neutrino flare" without observing quasi-simultaneous activity in the gamma-ray bands, posing challenges to established models. Electron-positron ($e^\pm$) pairs generated from the accretion disks have been amply proposed as a mechanism of bulk acceleration of sub-relativistic and relativistic jets. These pairs annihilate inside the source producing a line around the electron mass, which is blueshifted in the observed frame (on Earth) and redshifted in the frame of the dissipation region of the jet.  The redshifted photons in the dissipation region interact with accelerated protons,  producing high-energy neutrinos that contribute significantly to the diffuse neutrino flux in the $\sim$ 10 - 20 TeV energy range in connection with gamma-rays from the photo-pion process which can be detected by future MeV orbiting satellites.     Based on this phenomenological model, we can explain the ``neutrino flare"  reported in 2014 - 1015.
\end{abstract}




\section{Introduction}

The discovery of a diffuse neutrino flux in the energy range from TeV to PeV by the IceCube telescope has opened a new window in astrophysics at very high energies \citep[VHEs; ][]{2013Sci...342E...1I, 2014PhRvL.113j1101A, 2015arXiv151005223T, 2017ICRC...35..981K}.  Because of distinct configurations and strengths of magnetic fields among sources and Earth,  cosmic rays (CRs) themselves cannot supply accurate information where these were accelerated \citep[e.g., ][]{2007Sci...318..938P}.  A similar situation due to extragalactic background light (EBL) occurs when TeV photons are emitted from sources at high redshifts \citep[e.g.][]{2008A&A...487..837F}. Whereas magnetic fields deviate CRs, and EBL absorbs TeV photons, high-energy neutrinos can largely travel and reach Earth giving reliable spatial information about sources.\\
The IceCube collaboration reported on September 22, 2017, at 20:54:30.43 UTC, the detection of a neutrino-induced muon-track event called IceCube-170922A, which had an energy of $\sim$ 290 TeV \citep{2018Sci...361.1378I}. The reconstructed direction of this event from statistical and systematic effects was R.A.=$77^{\circ}.43^{+0.95}_{-0.65}$ and  Dec=$+5^{\circ}.72^{+0.50}_{-0.30}$ (J2000.0) which was consistent with the location of blazar TXS 0506+056 when it exhibited very high activity in gamma-rays, X-rays, optical and radio bands. Shortly after the detection,  IceCube analyzed the archival muon-neutrino data collected over 9.5 years in the direction of this blazar. This collaboration reported an excess of high-energy neutrinos over the atmospheric background between September 2014 and March 2015 \citep{2018Sci...361..147I}. This excess was associated with a large number of neutrinos ($13\pm 5$)  called ``neutrino flare", which had energies in the range of $\sim$ 10 - 20 TeV \citep{2019ApJ...874L..29R}. Surprisingly, this ``neutrino flare" was detected without significant activity in the electromagnetic bands.   This blazar was associated with a redshift of $z=0.3365\pm 0.0010$ \citep{2018ApJ...854L..32P}, after the identification of three weak emission lines ([O II] 327.7 mm, [0 III] 500.7 mm and [NII] 658.3 mm).\\
Several interpretations about the IceCube-170922A event and the ``neutrino flare"  with and without quasi-simultaneous gamma-ray activity have been suggested.   Taking into account the Fermi data, \cite{2018MNRAS.480..192P} claimed that during the period associated with the  ``neutrino flare",  the broadband spectral energy distribution (SED) of this blazar could have been hardened above 2 GeV.  However, \cite{2019arXiv190110806F} argued that this atypical behavior might not be relevant.  \cite{2018ApJ...865..124M} studied the IceCube-170922A event and the ``neutrino flare" from TXS 0506+056 and concluded that a two-zone synchrotron-self Compton (SSC) model could naturally explain both detections in the framework of the photomeson or the hadronuclear processes.  \cite{2019ApJ...881...46R} reconstructed the minimum target photon spectrum to explain the  ``neutrino flare", and the electromagnetic emission of the secondary particles. They concluded that the same process could not have originated neutrinos and gamma-rays.   \cite{2019NatAs...3...88G} demonstrated that the IceCube-170922A event could be explained with a moderate increase in the CRs during the flaring activity.   \cite{2019ApJ...874L..29R} investigated whether the models adopted to describe the flaring activity in 2017 could be used to interpret the ``neutrino flare" detected in 2014-2015.  They did not find any model that could describe simultaneously the ``neutrino flare" and the features of the flaring activity detected in 2017.\\ 
It has been ample proposed that blazar jets have a large amount of electron-positron ($e^\pm$) pairs generated from the accretion disks, which serve as a mechanism of bulk acceleration \citep{1982MNRAS.198.1109P, 1998Natur.395..457W, 1995ApJ...441...79B}. In particular cases,  \cite{1999MNRAS.305..181B} and \cite{2002ApJ...565..163I, 2004ApJ...601...78I} studied pair plasmas ejected with sub-relativistic and relativistic velocities, respectively.    \cite{1999MNRAS.305..181B} investigated the $e^\pm$ pairs created in gamma-gamma interactions outside the accretion disks. The author found that the pair plasma was ejected with mildly relativistic velocities, and the expected annihilation line produced by the pairs inside the source and observed on Earth was blueshifted and broadowing.    \cite{2002ApJ...565..163I, 2004ApJ...601...78I} considered a pair plasma from a Wein equilibrium state at relativistic temperatures.  They found that the pair plasma inside the outflow could be relativistically accelerated, and the emission from the photosphere could be detected as MeV-peaked flux. Later, \cite{2015APh....71....1F} computed the hadronic interaction between radiation coming from this plasma and protons accelerated in the dissipation region to explain orphan flares reported in some blazars \citep{2005ApJ...621..181D, 2011ApJ...738...25A}.\\
%
%
%
\\
In this paper, we present a phenomenological model in which photons produced by the pair annihilations at the base of the outflow reach the jet's dissipation region and interact with accelerated protons. We show that this model can explain a large diffuse neutrino flux in the 10 - 20 TeV energy range and the ``neutrino flare"  reported in 2014 - 2015.  We consider the pair plasma with the features described in  \cite{1999MNRAS.305..181B} and \cite{2002ApJ...565..163I, 2004ApJ...601...78I}.  We arrange the paper as follows. In Section 2, we introduce the one-zone lepto-hadronic scenario.  In Section 3, we compute the neutrino production from a sub-relativistic plasma scenario and a relativistic plasma in a Wein equilibrium state introduced by  \cite{1999MNRAS.305..181B} and \cite{2002ApJ...565..163I, 2004ApJ...601...78I}, respectively. Further,  we show the connection between MeV gamma-ray photons and the diffuse neutrino flux.  In Section 4, we apply our model, as a particular case, to interpret the  ``neutrino flare" reported in the source TXS 0506+056,  and finally, in Section 5, we present the discussion and summary.   We will use throughout the paper natural units $c=\hbar=1$ and  $Q_{\rm x}=Q/10^{\rm x}$ in c.g.s..  We consider the Hubble and the cosmological constants given by  $H_{\rm 0}=69.6\,{\rm km\,s^{-1}\,Mpc^{-1}}$,  $ \Omega_{\rm \Lambda}=0.714$ and $\Omega_{\rm m}=0.286$, respectively \citep{2016A&A...594A..13P}.  We use prime quantities in the rest frame of the black hole, and unprimed quantities in the observer frame and the comoving frame of the dissipation radius and the plasma. For instance,  $E'$ is the energy in the black hole rest frame, $E$ is the energy in the observer frame (on Earth), $\epsilon$ is the energy in the comoving frame of the dissipation radius and $\varepsilon$ is the energy in the comoving frame of the plasma.\\
\section{One-zone Lepto-hadronic scenario}\label{section2}
The most extensively accepted scenarios to interpret the  broadband SED of blazars  are:  i)  the one-zone SSC model in the leptonic scenario \citep[e.g.][]{2008ApJ...686..181F, 2011ApJ...736..131A, 2016ApJ...830...81F},  and ii) the one-zone proton synchrotron \citep{2001APh....15..121M, 2003APh....18..593M} and photo-hadronic interactions \citep[e.g.][]{2003ApJ...586...79A, ato01, 2013ApJ...768...54B, 2014MNRAS.441.1209F,  2015APh....70...54F,  2016ApJ...830...81F} in the hadronic scenario.  Irrespective of the scenario, we estimate the total kinetic luminosity of the jet through $$L_{\rm jet}=\sum_{i=e, p,  B} L_i\,,$$ where $L_e$, $L_B$ and $L_p$ are the luminosities due to electrons, magnetic field and protons, respectively, which depending on the source the luminosity associated to each set of particles or magnetic field could contribute more than another.\\
%
%
\subsection{Leptonic scenario}
We use a homogeneous one-zone model considering an electron population described with double-break power laws \citep{2011ApJ...736..131A}

\begin{equation}
\frac{dn_e}{d\gamma_e} =  N_{0,e}%
   \begin{cases}
     \gamma_e^{-\alpha_1}& \gamma_{\rm e, min}\leq \gamma_e\leq \gamma_{\rm c1}  \\
     \gamma_e^{-\alpha_2}   \gamma_{\rm c1}^{\alpha_2-\alpha_1}& \gamma_{\rm c1}<\gamma_e\leq \gamma_{\rm c2}  \\
     \gamma_e^{-\alpha_3}   \gamma_{\rm c1}^{\alpha_2-\alpha_1}\gamma_{\rm c2}^{\alpha_3-\alpha_2}& \gamma_{\rm c2}<\gamma_e\leq \gamma_{\rm e, max}
   \end{cases}
   \label{espsynm}
\end{equation}

with $N_{0,e}$  the number density of electrons, $\alpha_1$,  $\alpha_2$ and  $\alpha_3$ the spectral indexes, and $\gamma_{\rm e, min}$, $\gamma_{\rm c1}$, $\gamma_{\rm c2}$ and $\gamma_{\rm e, max}$ the electron Lorentz factors for minimum, breaks (1 and 2) and maximum, respectively.  This electron population is injected within the spherical dissipation  ``blob"  region of the jet ($r_{\rm b}$) which moves with a constant ultra-relativistic speed $\beta_b=\sqrt{1-1/\Gamma^2_b}$  in a collimated jet.  The term $\Gamma_b$ corresponds to the bulk Lorentz factor of the dissipation region.\\
The kinetic power ratio between the magnetic field and ultra-relativistic electrons can be estimated through the equipartition parameter $\lambda_{\rm B, e}=L_B/L_e$, where the luminosities carried by the magnetic field and electrons are \citep[e.g., see][]{2013ApJ...768...54B, 2014ApJ...783...44F,  2019ApJS..245...18F}
{\small
 \be
 L_B\simeq \frac18 r_b^2 \Gamma^2_b\, B^2\,,
 \ee
 }
and
{\small
 \be\label{Le}
 L_e\simeq \pi r_b^2 \Gamma^2_b m_e\int^{{\gamma_{\rm e,max}}}_{\gamma_{\rm e,min}} \gamma_e\,\frac{dn_e}{d\gamma_e}  d\gamma_e\,,
 \ee
 }
respectively, with $m_e$ the electron mass.\\
The electron population confined by the magnetic field ($B$) in the dissipation region begins to radiate photons by the synchrotron process and scatter them via inverse Compton scattering. \cite{2016ApJ...830...81F} and  \cite{2017ApJS..232....7F} explicitly illustrated and discussed the electron Lorentz factors, the timescales, and the synchrotron and inverse Compton scattering spectra with their spectral breaks. 
\subsection{Hadronic scenario}
Protons are co-accelerated with electrons and confined, in turn, in the magnetic field's dissipation region.  For this scenario, it is essential to define the proton luminosity as
{\small
 \be\label{Lp}
 L_p\simeq \pi r_b^2 \Gamma^2_b m_p\int^\infty_{1} \gamma_p\,\frac{dn_p}{d\gamma_p}  d\gamma_p\,,
 \ee
 }
where $m_p$ is the proton mass and $\gamma_{\rm p, max}$ is the maximum Lorentz factor for protons.   Protons are mainly cooled down by photo-hadronic interactions which encompass two mechanisms: photo-pion production and Bethe-Heitler (BH) pair production.\\
%
%
%
\subsubsection{Photo-pion production}
The proton energy loss rate due to photopion production is given by $p+\gamma \to  \pi^0 + p$ or $\pi^\pm +n (\Delta^{++})$. Then,  neutral pions decay into two photons ($\pi^0\to \gamma + \gamma$) and charge pions decay into neutrinos and anti-neutrinos  $\pi^+ (\pi^-)\to \mu^+(\mu^-) +\nu_\mu (\bar{\nu}_\mu)$  and $\mu^+ (\mu^-)\to e^+(e^-)+\nu_e (\bar{\nu}_e)+ \bar{\nu}_\mu (\nu_\mu)$.   The efficiency of the photo-pion production calculated through dynamical ($t_d\simeq \frac{r_b}{\Gamma_b}$) and photo-pion ($t_{p \pi}$) timescales in the comoving frame is given by \citep{ste68,wax97}
\begin{equation}\label{fp}
f_{p\pi}\simeq \frac{t_d}{t_{p \pi}}  = \frac{r_b}{2\Gamma_{\rm b}\gamma_p^2} \int^\infty_{\bar{\epsilon}_{th}}  d\bar{\epsilon} \sigma_{p\pi}(\bar{\epsilon}) \kappa_p(\bar{\epsilon}) \bar{\epsilon} \int^\infty_{\bar{\epsilon}/2\gamma_p} d\epsilon_\gamma \frac{n(\epsilon_\gamma)}{\epsilon^2_\gamma}\,,
\end{equation}
where $\sigma_{p\pi}  \simeq \sigma_{\rm pk}\, \xi_{p\gamma}\,\frac{\Delta\epsilon_{pk}\,}{\epsilon_{pk}}$ with $\sigma_{\rm pk}\approx 5\times\,10^{-28}\,\, {\rm cm^2 }$ is the photo-pion cross section,   $\Delta\epsilon_{\rm pk}$=0.2 GeV, $\epsilon_{\rm pk}\simeq$ 0.3 GeV, $\xi_{p\gamma}\simeq 0.2$ and $n(\epsilon_\gamma)$ is the photon spectrum.  We consider a mono-energetic photon distribution 
\begin{equation}\label{mono}
 n(\epsilon_\gamma)=  \frac{u_\gamma}{\epsilon_\gamma} \delta(\epsilon-\epsilon_\gamma)\,,
\end{equation}
and the energy density of target photons given by
\begin{equation}
 u_\gamma \approx \frac{L_{\rm \gamma, ph}}{4 \pi\,\Gamma_b^2\, r_b^2}\,,
\end{equation}
with $L_{\rm \gamma, ph}$ the luminosity of the seed photons. The efficiency of the photo-pion production can be written as 
\begin{equation}\label{f_ppi}
f_{p\pi} \approx 2 \frac{u_\gamma}{\epsilon_\gamma} \sigma_{pk}\, r_b\, \epsilon_{\rm pk}\,\left[1-\left(\frac{E'^{th}_p}{E'_p}\right)^2\right]\,,
\end{equation}
where $E'^{th}_p$ corresponds to the threshold of the proton energy.
\subsubsection{Bethe-Heitle pair production}
The proton energy loss rate due to BH pair production is given by $p+\gamma\to p+e^++e^-$.   The efficiency of the BH pair production calculated through the dynamical and BH ($t_{pe}$) timescales is given by \citep{2015MNRAS.447...36P}
\begin{equation}
f_{pe}\simeq \frac{t_d}{t_{pe}}  = \frac{3}{8 \pi \gamma_p} \sigma_T  \alpha_f \frac{m_e}{m_p} \int^\infty_{2}  d\kappa \; n \left( \frac{\kappa}{2\gamma_p} \right) \frac{\phi(\kappa)}{\kappa^2}\,,
\end{equation}
where $\alpha_f$ is the fine structure constant, $\kappa=2\gamma_p \epsilon_\gamma/m_e$, $\sigma_T=6.65\times 10^{-25}\,{\rm cm}^2$ is the Thompson cross section,  and $\phi(\kappa)$ is a function defined in \cite{1992ApJ...400..181C}.  Considering the mono-energetic photon distribution given by eq. (\ref{mono}), the efficiency of the BH pair production can be written as
\begin{equation}\label{f_pe}
f_{pe} \approx  \frac{3}{4 \pi} \sigma_{T} \alpha_f \frac{m_e}{m_p}\frac{u_\gamma}{\epsilon_\gamma} \frac{\phi\left( \kappa \right) }{\kappa^2} r_b\,.
\end{equation}
\subsubsection{Secondary pair productions}
Synchrotron emission is expected from secondary pairs created in photo-pion and BH pair production. The characteristic photon energy radiated by synchrotron is $\epsilon_{\gamma}=\frac{q_e}{2\pi\,m^3_e} B \epsilon^2_{e, i}$, where $q_e$ is the elementary charge and $\epsilon_{e, i}$ is the energy of the secondary pairs with $i=p\pi$ and $pe$ for  photo-pion and BH pair production, respectively.
\paragraph{The secondary pairs from photo-pion process.}
The electron energy generated by the photo-pion production is
\begin{equation}
 \epsilon_{e,p\pi} \simeq \frac{1}{4} \kappa_p \epsilon^{\rm th}_{p}\,.
\end{equation}
The synchrotron spectrum estimated through the proton spectrum and the efficiency of the photo-pion production can be written as \citep{2015MNRAS.447...36P}
\be
L_{\rm E'_{syn, p\pi}}=\frac18 f_{p\pi}\,L_{E'_p}\,,
\ee
with $f_{p\pi}$ given by eq. (\ref{f_ppi}).
\paragraph{The secondary pairs from BH process.}
The electron energy generated by the HE pair  production is \citep{PhysRevD.78.034013}
\begin{equation}
\epsilon_{\rm e,pe}=\frac{\gamma_p}{1+4 \gamma_p \epsilon_\gamma / m_p} \left( \sqrt{\gamma_p \epsilon_\gamma} + \sqrt{\gamma_p \epsilon_\gamma -m_e}\right)^2\,.
\end{equation}
The synchrotron spectrum estimated through the proton spectrum and the efficiency of the photo-pion production can be written as \citep{2015MNRAS.447...36P}
\be
L_{\rm E'_{syn, pe}}=f_{pe}\,L_{E'_p}\,,
\ee
with $f_{pe}$ given by eq. (\ref{f_pe}).
\section{Neutrino Production}
In the scenario of the photo-pion and BH processes, the $\pi^\pm$ decay products generate neutrinos and anti-neutrinos.  We calculate the efficiencies of the photo-pion and BH processes from eqs. (\ref{f_ppi}) and (\ref{f_pe}), respectively.  By comparing the acceleration, the cooling, and the dynamical timescales, we can estimate the maximum energy that protons can reach in the dissipation region.   The acceleration timescale for protons with energy $\epsilon_p$ is characterized by  $t_{\rm acc}=\eta_{\rm L}\epsilon_p/q_eB$ where $\eta_{\rm L}$ is order of unity  \citep[e.g.][]{2012ApJ...753...40F,2019JCAP...08..023F}.\\
The maximum of the neutrino spectrum calculated through the photo-pion production can be obtained through the proton spectrum given by \citep{2014PhRvD..90b3007M}
\bary\label{pg}
E'_{\rm \nu} L_{E'_\nu}\simeq\frac{3}{8} f_{p\pi} E'_{\rm p} L_{E'_{\rm p}}\,,
\eary
where $f_{p\pi}$ is the corresponding efficiency.  The characteristic energy can be roughly estimated as  
\be\label{nu_br}
E'_{\rm \nu,b}\approx 51.3\,{\rm GeV} \, \Gamma_{b} \Gamma_{\rm rel}    \left(\frac{\epsilon_{\rm \gamma}}{511\,{\rm keV}}\right)^{-1}\,,
\ee
where $\epsilon_{\rm \gamma}$ is the energy of the seed photons corresponding to the pair annihilations given in the sub-relativistic and relativistic pair plasma scenarios (see the following subsections), and $\Gamma_{\rm rel}$ is the relative Lorentz factor between the pair plasma where the photons emerge and the dissipation region (see eq. \ref{gamma_rel}).
\subsection{A sub-relativistic pair plasma scenario}
In the optically thin outflow,  pairs get away without annihilation, and the corresponding luminosity in the electron-positron rest mass is estimated through the number of high-energy photons interacting above the disk.    In the optically think case, pairs  annihilate before they can escape from the source ($\frac{t_{\rm an}}{t_{\rm esc}}\simeq \frac{1}{n_{\pm}\sigma_T\,r}<1$), with  $n_{\pm}$ the density of the pairs.  These pairs are in Compton equilibrium with radiation at a temperature of  $\sim 10\,{\rm keV}$.   The plasma at the base of the outflow has a bulk velocity near the equilibrium $\beta_{\rm p}\sim 0.3$ incrementing to $\beta_{\rm p}\sim 0.7$ at the photosphere \citep{1999MNRAS.305..181B}.\\
The escaping photons can be detected by a distant observer as an annihilation line of width equivalent $\sim 10\, {\rm keV}$.\\
The observed luminosity in photons with energy (in the observer frame)
\be\label{E_line}
E_\gamma\approx  \frac{1}{1+z} \frac{m_e}{\Gamma_{\rm p}(1-\beta_{\rm p} \cos\theta)  }
\ee
corresponds to a photon density $n\simeq\frac{L_{\rm \gamma,ph}}{\pi r^2\,m_e}$.  The term $\theta$ is an arbitrary angle with respect to the line between the source and observer  and $\Gamma_{\rm p}=1/\sqrt{1 - \beta^2_{\rm p}}$ is the bulk Lorentz factor of the pair plasma.   The optical depth for interacting photons is $\tau_{\gamma\gamma}\simeq n\sigma_{\gamma\gamma}\,r$, where $\sigma_{\gamma\gamma}$ is the average cross section for photon interactions.\\
\\
In the transition zone from the optically thick to thin $e^\pm$ envelope, pairs are  $n_\pm\sim (\sigma_T\,r_{\rm ph})^{-1}$ with $r_{\rm ph}$ the photosphere radius. The escaping pairs can be estimated as $F_{\pm}\sim\frac{1}{\sigma_T\,r_{\rm ph}}$ and the corresponding emerging pair luminosity as $L_{\pm}=2\pi\,r^2_{\rm ph}\,F_{\pm}$. Numerical calculations of the density profile of the pair outflow generated above a disk with different parameters for an optically thick and thin outflow are given in \cite{1999MNRAS.305..181B}.
\subsection{A relativistic plasma in Wein equilibrium state}
In this scenario, the jet's base connects the black hole with the Wein fireball. At the initial state,  e$^\pm$ pairs in quasi-thermal equilibrium inside the initial scale $r_o=2r_g=4GM$ form the Wein fireball,  being  $M$ the black hole mass, $r_g$ the gravitational radius and $G$ the gravitational constant.  In the first state, photons inside the Wein fireball are at relativistic temperature defined through microscopic processes. The internal energy transforms into kinetic energy, and the Wein fireball expands by its radiation pressure.   As a result of this expansion, the temperature decreases, and the bulk Lorentz factor increases in the first state.  The initial optical depth is \citep{2002ApJ...565..163I,2004ApJ...601...78I}
\be\label{opt}
\tau_o \simeq \frac{n_{e,o}\,\sigma_T\,r_{o}}{\Gamma_{o}}\,,
\ee
where $\Gamma_{o}=1/\sqrt{1 - \beta^2_{\rm o}}$ is the initial Lorentz factor of the plasma and $n_{e,o}$ is the initial electron density which is given by
\be\label{elec_den}
n_{e,o}=\frac{1}{4\sigma_TGM} \frac{1}{\Gamma^2_{o}\,\beta_{o} \langle \gamma_{e,o}\rangle} \left(\frac{m_p}{m_e}\right) \left(\frac{r_g}{r_o}\right)^2\left(\frac{L_j}{L_{Edd}}\right)\,, 
\ee
where $L_j$ is the total luminosity of the jet, $L_{Edd}=2\pi m_pr_g/\sigma_T$ is the Eddington luminosity, $\langle\gamma_{e,o}\rangle=K_3(1/\theta_o)/K_2(1/\theta_o)-\theta_o$ is the average Lorentz factor of electron thermal velocity, $\theta_o=T_o/m_e$ is the initial temperature normalized to electron mass  and $K_i$ is the modified Bessel function of integral order.  By considering  the conservation equations of energy  and momentum for a steady and spherical flow \citep{2004ApJ...601...78I,2002ApJ...565..163I}, during the expansion of the Wein fireball the number density of photons and pairs evolve as $n_\gamma+2n_e=3n_{\rm e,0}\,\left(\frac{r_0}{r} \right)^3$,  the temperature as $\theta=\theta_0\frac{r_0}{r}$,   the bulk Lorentz factor as $\Gamma_{\rm p}=1/\sqrt{1 - \beta^2_{\rm p}}=\Gamma_{0}\frac{r}{r_0}$ and the optical thickness as $\tau\equiv \tau_0\left(\frac{r}{r_0} \right)^{-3}$.  Finally, the quasi-thermal radiation scape at the photosphere (defined at $\tau=1$), for a radius of $r_{\rm ph}\simeq\tau_0^\frac13 r_0$.   The numerical results of the relevant quantities at the photosphere such as the radius, the temperature and the velocity are $2.7\lesssim \frac{r_{\rm ph}}{r_{\rm g}}\lesssim 9.9$,  $0.2\lesssim \theta_{\rm ph}\lesssim1.5$,  $1.3\lesssim (\Gamma_{\rm p} \beta_{\rm p})_{\rm ph} \lesssim5.8$ and $0.1\lesssim \frac{L_{\rm j}}{L_{\rm Edd}} \lesssim25$, respectively  \citep{2004ApJ...601...78I}.
\subsection{Interactions between photons from the pair plasma and protons within the dissipation region}
\begin{figure}
{\centering
\resizebox*{0.44\textwidth}{0.27\textheight}
{\includegraphics{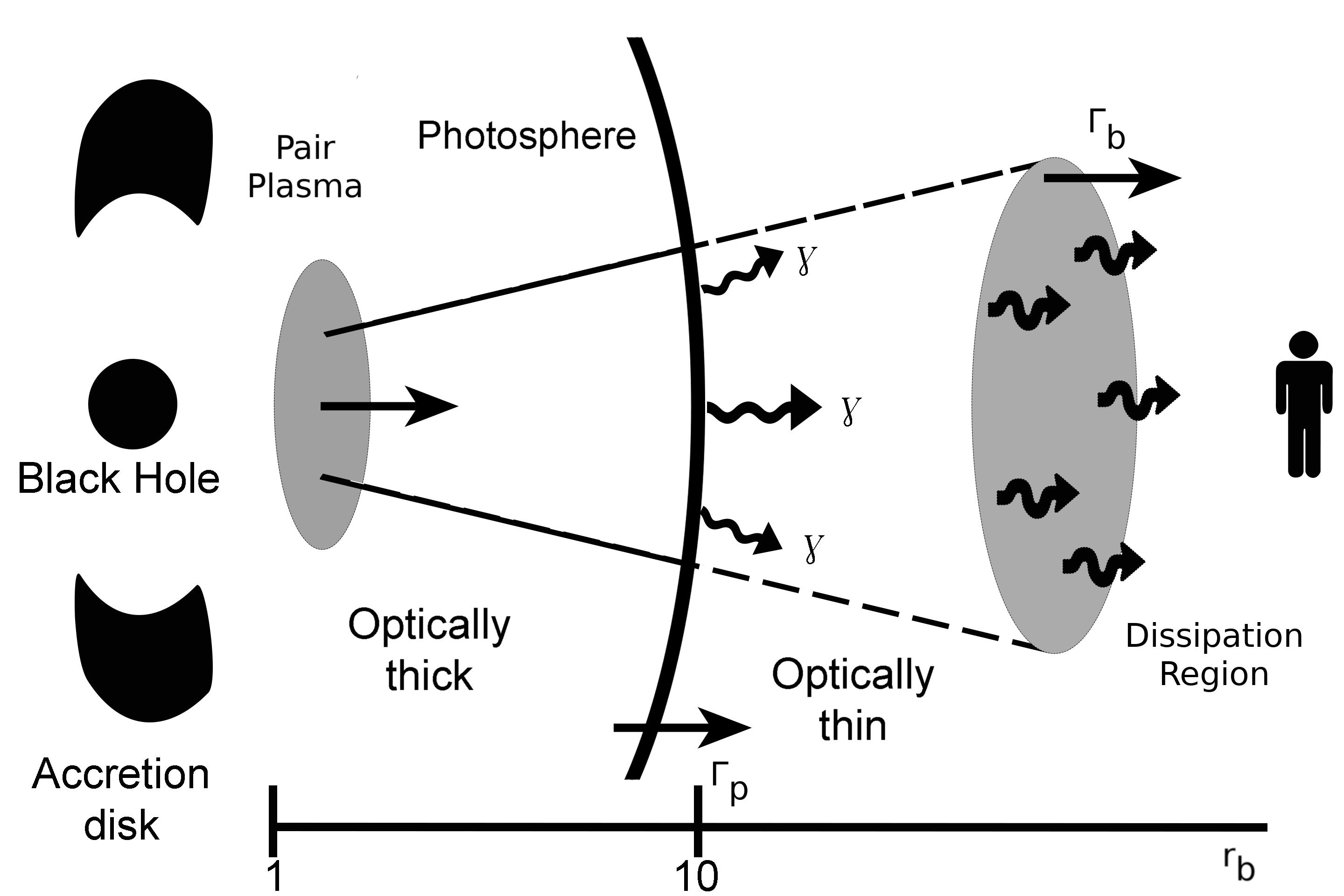}}
}
\caption{Schematic representation of the dynamics of the pair plasma  and the dissipation region of TXS 0506+056. Physical distances and radii are not to scale.}
\label{squeme1}
\end{figure}

Figure \ref{squeme1} shows a schematic representation of the photo-hadronic interactions between the radiation generated inside the pair plasma and the accelerated proton in the dissipation region.  We consider as seed photons those from the annihilation line released from the photosphere.  As follows, we describe the dynamics, and Lorentz factors evolved in each frame.\\
Emerging photons generated in the pair plasma and released from the photosphere will have energies given by eq. (\ref{E_line}).  These are observed on Earth as an annihilation blueshifted line with an equivalent width of some keV \citep{1999MNRAS.305..181B, 2002ApJ...565..163I, 2016Natur.531..341S}. 
Emerging photons from the plasma are redshifted in the frame of the dissipation region, which moves with a Lorentz factor $\Gamma_b$.    The relative Lorentz factor  between the pair plasma and the dissipation region is

\be\label{gamma_rel}
\Gamma_{\rm rel}=\Gamma_{\rm b}\Gamma_{\rm p}(1 - \beta_{\rm b}\beta_{\rm p})\,.
\ee
In simple terms, the annihilation line is redshifted in the frame of the dissipation radius.  For instance,  considering a typical value of Lorentz factor for blazars $\Gamma_{\rm b}=10$ \citep{2014PhRvD..90b3007M} and a sub-relativistic velocity with $\beta_{\rm p}=0.3$ \citep{1999MNRAS.305..181B}, the relative bulk Lorentz factor becomes $\Gamma_{\rm rel}\simeq7.5$.  Therefore, in the jet's comoving frame, the annihilation line has an energy of dozens of keV corresponding to
\be\label{seed_Ene}
\epsilon_\gamma\approx 25.1\,{\rm keV}\,  \Gamma^{-1}_{\rm rel,1}  \left(\frac{\varepsilon_{\rm \gamma}}{511\,{\rm keV}}\right)\,.
\ee
Photons from the photosphere reach the  dissipation region and interact with the accelerating protons producing neutrinos.  The characteristic energy can be roughly estimated as  
\be\label{nu_br}
E'_{\rm \nu,b}\approx 5.1\,{\rm TeV} \, \Gamma_{b, 1} \Gamma_{\rm rel, 1}    \left(\frac{\epsilon_{\rm \gamma}}{511\,{\rm keV}}\right)^{-1}\,.
\ee
If we consider a relativistic plasma  with velocity  $\beta_{\rm p}=0.98$, then the Lorentz factor becomes $\Gamma_{\rm rel}\simeq1.2$. In this case,  the annihilation line would have an energy of $\epsilon_\gamma\approx 0.3\,{\rm MeV}\,$  and the characteristic neutrino energy becomes $E'_{\rm \nu,b}\approx 0.6\,{\rm TeV}$.\\
\vspace{1cm}
\subsection{Connection between MeV gamma-ray photons and diffuse neutrino flux}
The neutrino and the gamma-ray spectra  from extragalactic BL Lac objects can be estimated through the expression
\bary
\phi(E_{\rm x}) &=& \frac{1}{4\pi \, H_0 } \int^{z_{\rm max}} \frac{dz }{ (1+z)^2 \sqrt{\Omega_m (1+z)^3 + \Omega_\Lambda} }\cr
&&\hspace{1.5cm} \times \,\int  dL_{\rm \gamma} \,\, \rho(z,L_{\rm \gamma}) \frac{L_{E'_{\rm x}}}{E'_{\rm x}}\,,
\eary
where $\rho(z,L_\gamma)$ is the gamma-ray luminosity function of BL Lac objects (per comoving volume in the range of  $\rm[log\,L_{\rm \gamma},\,log\,L_{\rm \gamma} + dlog\,L_{\rm \gamma}]$ at a  redshift $z<z_{\rm max}$) and $E'_{\rm x}$ is the energy of neutrinos ($x=\nu$) and synchrotron due to secondary pairs ($x=e^\pm$).  We consider the gamma-ray luminosity function reported in \cite{2014MNRAS.441.1760Z}.\\
In order to compute the individual contribution of each BL Lac, we assume a proton distribution described by

\begin{equation}
\frac{dn_p}{d\gamma_p} =A_{\rm p}\gamma_{\rm p}^{-\alpha_p}\,,
\end{equation}

with $\alpha_{\rm p} \approx 2$ the spectral proton index and $A_p$ the normalization constant. This normalization is estimated assuming that the proton luminosity corresponds to a fraction $\eta_{\rm p}$ of the Eddington luminosity (i.e. $L_{\rm p} = \eta_{\rm p} L_{\rm edd}$ ).

\begin{figure}
{\centering
\resizebox*{0.5\textwidth}{0.3\textheight}
{\includegraphics{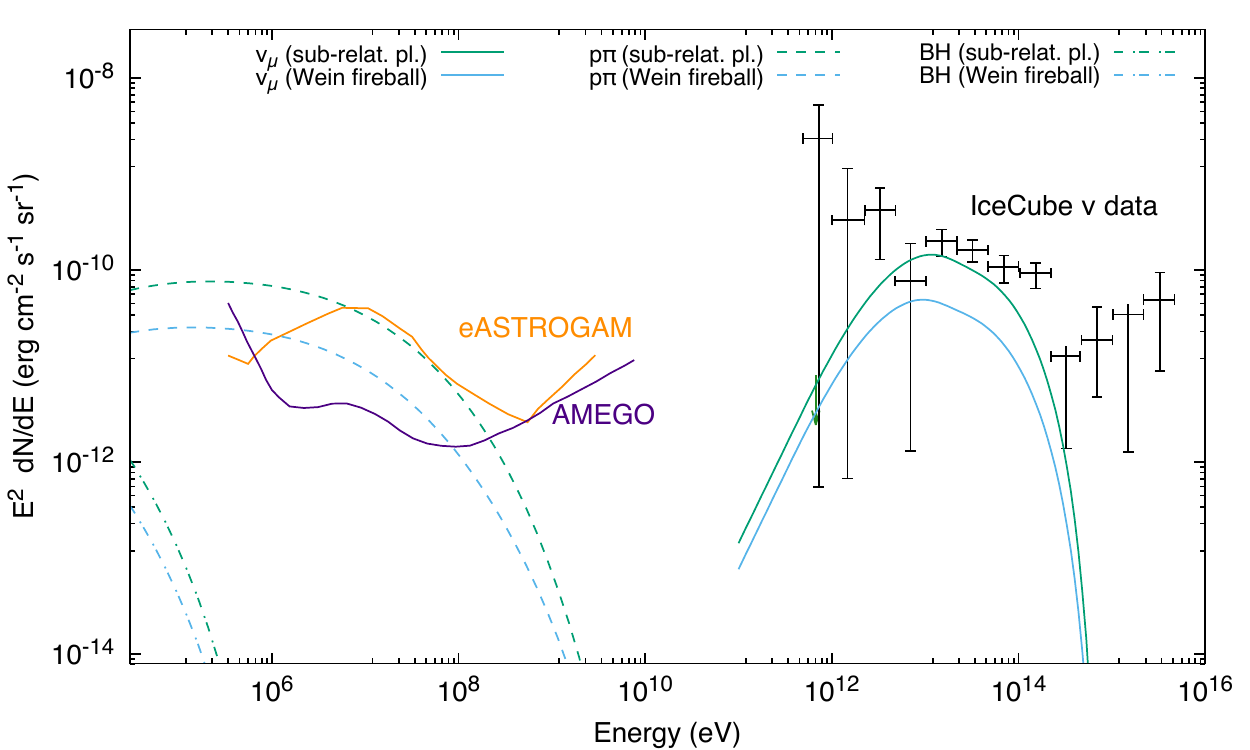}}
}
\caption{We show The neutrino and gamma-ray spectra from a sub-relativistic pair plasma scenario and a relativistic plasma in Wein equilibrium state. The sensitivities of eASTROGAM and AMEGO  experiments are taken from \citep{2018JHEAp..19....1D} and \citep{2019arXiv190707558M}, respectively.  These sensitivities are given in ${\rm GeV\,cm^{-2}\,s^{-1}}$.   We take the IceCube data points from \citep{2017arXiv171001191I}. }  
\label{Nev}
\end{figure}

The neutrino and gamma-ray synchrotron fluxes from a single source are calculated following the hadronic model described in section \ref{section2}.   We consider the seed photons as those created inside a sub-relativistic pair plasma scenario and a relativistic plasma in the Wein equilibrium state.  In both scenarios, we describe the seed photons with the monoenergetic distribution function given in eq. (\ref{mono}).  Further, we assume that the gamma-ray luminosities radiated from the photosphere and the dissipation region are similar.  The parameter values  used in Figure \ref{Nev} are reported in Table \ref{table1}. \\
Figure \ref{Nev} shows the neutrino and gamma-ray synchrotron spectra from extragalactic BL Lac objects with redshifts $z<2$. We show the neutrino and gamma-ray spectra from a sub-relativistic pair plasma scenario and a relativistic plasma in Wein equilibrium state.   The neutrino flux from the sub-relativistic pair plasma scenario is larger than the relativistic plasma and significantly contributes to the 1 - 30 TeV energy range.   The gamma-ray synchrotron flux at a few MeV could be observed in eASTROGAM from the sub-relativistic pair plasma scenario but not from the relativistic plasma. The gamma-ray fluxes from the secondaries pairs in both scenarios could be observed in the AMEGO experiment.
\vspace{0.5cm}
\section{Particular Case: TXS 0506+056}
\subsection{Multiwavelength and neutrino data}
We retrieve the Fermi-LAT (Large Area Telescope) data  using the Fermi public database.\footnote{http://fermi.gsfc.nasa.gov/ssc/data}  We report the Fermi-LAT gamma-ray fluxes between the energy range 0.1 - 300 GeV. The Swift-BAT (Burst Area Telescope),   XRT (X-ray Telescope) and  UVOT (Ultra Violet/Optical Telescope) data used in this work are  publicly available.\footnote{http://swift.gsfc.nasa.gov/cgi-bin/sdc/ql?}  The Owens Valley Radio Observatory \citep[OVRO;][]{2011ApJS..194...29R}  data are publicly available.\footnote{http://www.astro.caltech.edu/ovroblazars/}  All-Sky Automated Survey for Supernovae (ASAS-SN) optical data used in this work are  publicly available.\footnote{https://asas-sn.osu.edu/variables} Archival data used in this work are publicly available. \\ 
Neutrino data collected with the IceCube collaboration,  VHE data collected with HESS (High Energy Stereoscopic System), Major Atmospheric Gamma Imaging Cherenkov (MAGIC),   Very Energetic Radiation Imaging Telescope Array System (VERITAS)  and  HAWC (High-Altitude Water Cherenkov) observatory,  gamma-ray data collected with the AGILE satellite, X-ray data collected with Nuclear Spectroscopic Telescope Array (NuSTAR) and INTErnational Gamma-Ray Astrophysics Laboratory (INTEGRAL),  optical data with Kiso (G-band), Kanata (R-band) and  SARA (UA) ground Telescopes and radio data collected with Very Large Area (VLA;  11 GHz)  are taken from \cite{2018Sci...361..147I}.
\subsection{Electromagnetic flaring activity in 2017}
On September 22, 2017, the IceCube-170922A event triggered the IceCube experiment \citep{2018Sci...361..147I}.  One week later, the Fermi-LAT collaboration announced high spatial coincidence activity with the blazar TXS 0506+056 and the IceCube-170922A event.  Immediately after that, this source was followed up by a multiwavelength campaign covering a wide range of the electromagnetic spectrum. For instance, the MAGIC telescopes monitored this source and detected VHE gamma-ray emission. \cite{2018Sci...361..147I} concluded that the blazar jet might accelerate CRs up to energies of several PeV. \\
Figure \ref{lightcurve} shows the multiwavelength (from radio to VHE gamma-ray) light curves of blazar TXS 0506+056 collected from 2008 August 22 to 2017 December 12.  The red dashed line shows the IceCube-170922A event. We show a blow-up of the flaring activity exhibited during the IceCube-170922A event in the gamma-ray, X-ray, optical, and radio bands.   Figure \ref{sed} shows the broadband SED of the blazar TXS 0506+056 with different curves corresponding to distinct models and contributions. We show the multiwavelength data collected during the flaring activity in salmon color, and the observed flux associated with the IceCube-170922A event for 7.5 (uncertainty in solid line) and 0.5 (uncertainty in dashed line) years in black points.  The triple-dotted-dashed black line corresponds to the best-fit curve proposed by \cite{2019NatAs...3...88G}. They considered CRs in the jet of   TXS 0506+056 and demonstrated that a moderate increase in the CRs during the flaring activity could yield a powerful increase of the neutrino flux with a range of blazar parameters. Several theoretical models have been proposed to interpret the IceCube-170922A event \citep[e.g. see][]{2018MNRAS.480..192P, 2018ApJ...865..124M, 2019NatAs...3...88G, 2019ApJ...881...46R, 2019arXiv190110806F,2019MNRAS.483L.127R, 2019A&A...630A.103B, 2018ApJ...864...84K, 2020ApJ...889..118Z}.
\vspace{1cm}
\subsection{Neutrino flare during 2014-2015}
\begin{figure}
{\centering
\resizebox*{0.5\textwidth}{0.4\textheight}
{\includegraphics{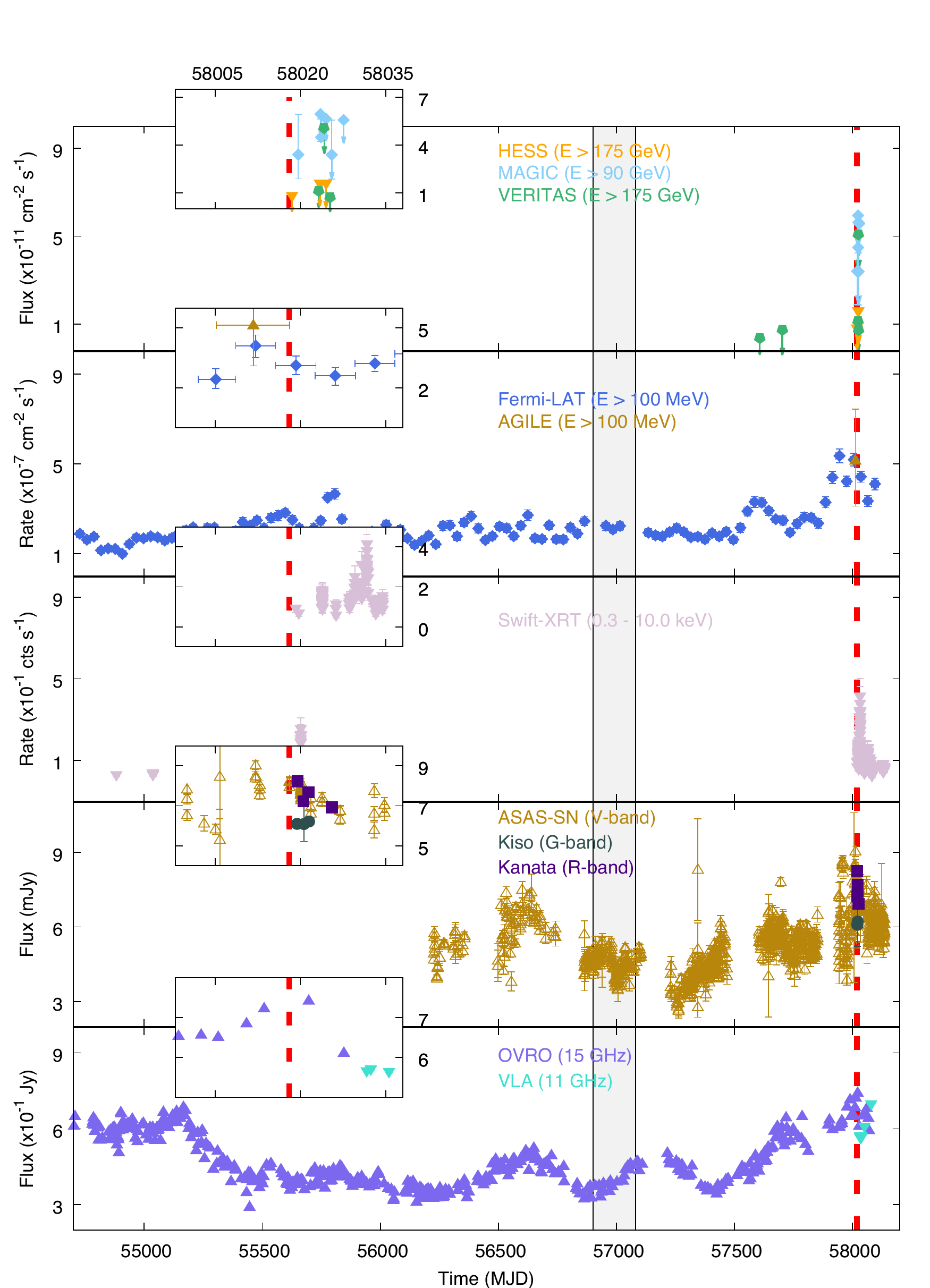}}
}   
\caption{TXS 0506+056 light curves are shown between 2008 August 22 to 2017 December 12, obtained with multiple orbiting satellites and ground-based observatories. We present from top to bottom: VHE, MeV - GeV $\gamma$-ray, X-ray, optical, and radio wavelengths.  The red dashed line shows the IceCube-170922A event, and the shadow area corresponds to the period associated with the ``neutrino flare". We show a blow-up of the flaring activity around the IceCube-170922A event in each electromagnetic band.}
\label{lightcurve}
\end{figure}

\begin{figure}
{\centering
\resizebox*{0.45\textwidth}{0.3\textheight}
{\includegraphics{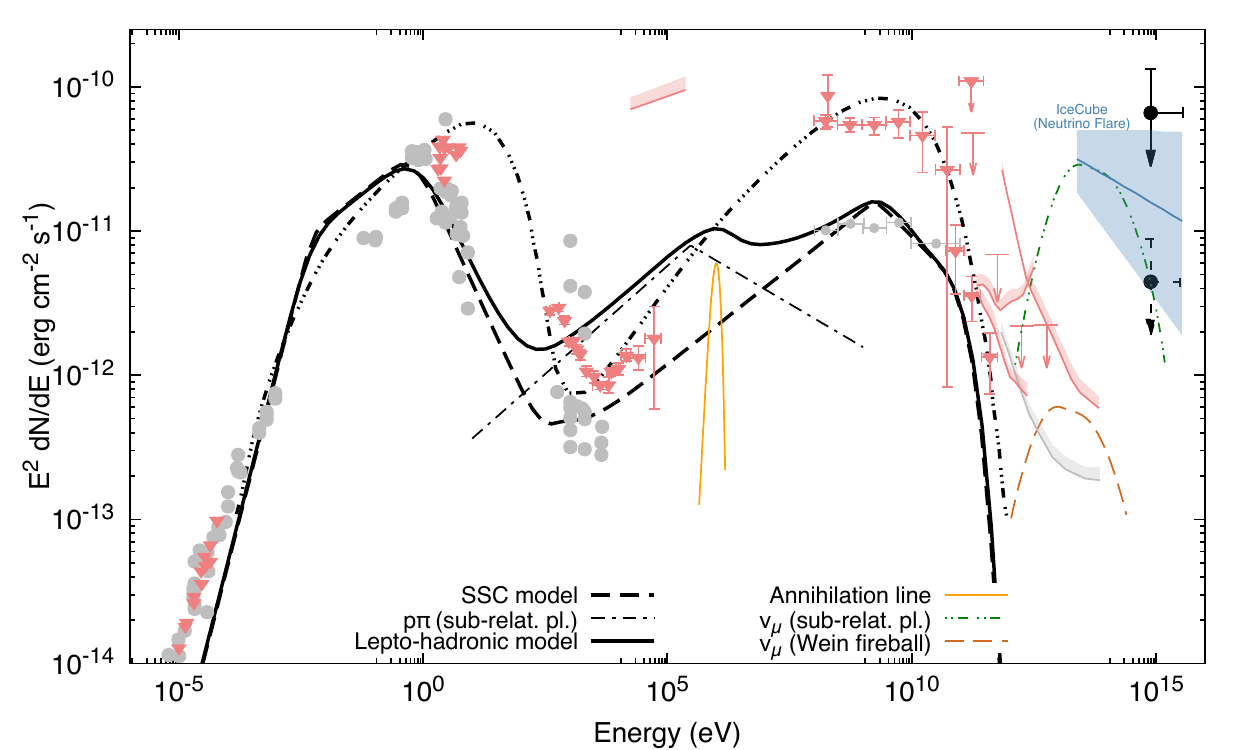}}
}   
\caption{We present the broadband SED of TXS 0506+056 with the model curves for lepto-hadronic models introduced in this work and \citep{2019NatAs...3...88G}.   The archival data points shown in gray correspond to the data before the flare in 2017, and the salmon data points correspond to the flare.  The triple-dotted-dashed black line corresponds to the best-fit curve proposed by \citep{2019NatAs...3...88G} and the solid black line to the lepto-hadronic used in this work. The solid yellow line represents the observed seed photons.  The double-dotted-dashed green line and the dashed brown line correspond to the neutrino spectra from the sub-relativistic pair plasma and the relativistic plasma in Wein equilibrium state, respectively. The observed flux associated with the IceCube-170922A event (black data point) is shown for 7.5 (uncertainty in solid line) and 0.5 (uncertainty in dashed line) years. The shadow region in blue corresponds to the range of energy and flux of the neutrino flare reported by IceCube Collaboration \citep{2018Sci...361..147I}.} %
\label{sed}
\end{figure}

The IceCube collaboration reported an excess of high-energy neutrinos concerning the atmospheric background between September 2014 and March 2015.  This excess called  ``neutrino flare" was associated with a large number of neutrinos  ($13\pm 5$) with energies in the range of $\sim$ 10 - 20 TeV \citep{2018Sci...361..147I, 2018MNRAS.480..192P}.     The shadow region in Figure \ref{lightcurve} corresponds to the fraction of the multiwavelength light curves associated with the timescale during which the ``neutrino flare" was observed.   Data points in gray color exhibited in Figure \ref{sed} correspond to the archival data.    These data show an intense flux in the soft X-rays at $\sim$ 1 - 2 keV, which was not collected during the observation of the ``neutrino flare" (see Figure \ref{lightcurve}).   Contrary to the flare observed in 2017, the gamma-ray, optical and radio fluxes were not detected in high activity during this period, suggesting different origins.\\
In order to describe the best-fit curve (solid black line) that describes the broadband SED in TXS 0506+056 (the archival data points), we use the SSC model presented in \cite{2016ApJ...830...81F, 2017ApJS..232....7F}.  We use the Chi-square $\chi^2$ minimization method implemented in the ROOT software package \citep{1997NIMPA.389...81B} and the procedure to obtain the values of the bulk Lorentz factor of the jet, the size of dissipation region,  the electron density, the strength of the magnetic field and the spectral indexes \citep{2016ApJ...830...81F}. We report in Table \ref{table2} the fitting and derived parameters. For instance, the values of fitting parameters of the bulk Lorentz factor,  the size of dissipation region and the strength of the magnetic field are in the range considered for other models \citep[e.g., see][]{2018ApJ...865..124M, 2019ApJ...874L..29R, 2019ApJ...881...46R}.\\  
The value of the minimum electron Lorentz factor ($\gamma_{\rm e,min}=8\times 10^2$) used in our model to fit the broadband SED indicates that the electron population is efficiently accelerated above the corresponding energy. Below this one, a distinct process that generates a hard electron spectrum accelerates these electrons.   The value of the electron luminosity derived is  $\approx 2$ times smaller than the Eddington luminosity $L_{\rm Edd}=3.8\times 10^{46}\,{\rm erg\,s^{-1}}$ which is estimated considering a black hole mass of $3\times 10^8\,M_{\odot}$ \citep{2019MNRAS.484L.104P}.   Using the value of the BH mass, the gravitational radius is $r_{\rm g}=8.9\times 10^{13}\,{\rm cm}$, which is two orders of magnitude smaller than the dissipation radius.   Given the best-fit values of the bulk Lorentz factor and the dissipation radius,  the variability timescale becomes $t_\nu=0.9\pm0.1\,{\rm days}$.  The value of the magnetic and electron luminosity ratio $\lambda_{\rm B, e}\approx0.03$ suggests that a principle of equipartition is not present in the jet of TXS  0506+056.   We find that the best-fit value of the second break of the electron distribution ($\gamma_{e,c2}$)  is due to synchrotron cooling, and the first break ($\gamma_{e,c1}$) due to the acceleration process.   Therefore, this model requires that the electron distribution below the second break be accelerated less efficiently.\\    
Based on the IceCube-170922A event \citep{2018Sci...361..147I} and the theoretical models used to interpret it \citep[e.g. see][]{2018MNRAS.480..192P, 2018ApJ...865..124M, 2019NatAs...3...88G, 2019ApJ...881...46R},  we also assume the existence of protons inside the dissipation region of jet.  Given the value derived of the maximum electron Lorentz factor, it is possible to estimate the maximum Lorentz factor for protons $\gamma_{\rm p, max}=\frac{m_p}{m_e}\gamma_{\rm e, max}=2.13\times 10^{11}$.  Although the Hillas condition is too optimistic,  the maximum energy that CRs could reach in the dissipation radius is $E_{\rm p, max}=eZ\,r_d\,B\,\Gamma \approx 3.46\times 10^{19}\,{\rm eV}$ for Z=1 \citep{1984ARA&A..22..425H}. 
Taking into account the best-fit value of the bulk Lorentz factor $\Gamma_{\rm b}=20$ and the numerical results about the velocity range of the $e^\pm$ outflow $0.3\lesssim\beta_p\lesssim0.7$ \citep{1999MNRAS.305..181B},  from eqs. (\ref{gamma_rel}) and (\ref{seed_Ene}),  the relative Lorentz factor and the energy of the annihilation line lie in the range of $9\lesssim \Gamma_{\rm rel} \lesssim 22$ and $15\lesssim \epsilon_\gamma \lesssim 25\,{\rm keV}$, respectively.    Protons interacting with these seed photons produce neutrinos with the characteristic energy in the range of $10 \lesssim E_{\rm \nu,b} \lesssim 20\,{\rm TeV}$ which corresponds to the range of events observed in the IceCube neutrino telescope.  The double-dotted-dashed green line and the dashed brown line correspond to the neutrino spectra estimated from the sub-relativistic pair plasma and the relativistic plasma in the Wein equilibrium state.  We can observe that the neutrino spectrum from the sub-relativistic pair plasma peaks at $\sim 3\times 10^{-11}\,{\rm erg\,cm^{-2}\,s^{-1}}$ and lies in the range (the shadow region in blue) reported by IceCube Collaboration  \citep{2018Sci...361..147I}.     The solid curve in yellow exhibited in Figure \ref{sed} corresponds to the seed-photon flux emitted from the pair plasma.    In addition to the 10 - 20 TeV neutrinos, VHE photons in the range of  $\sim$ 20 - 40 TeV and secondary pairs from photo-pion and BH processes are created. The respective efficiencies of the photo-pion and BH processes are  $f_{p\pi}\approx1$ and $f_{pe}\approx 0.06$, respectively.   The synchrotron energy break  and the maximum flux of secondary pairs are  $\approx 0.1\,{\rm MeV}$ and $\approx 9\times 10^{-12}\,{\rm erg\,cm^{-2}\,s^{-1}}$ for photo-pion process, and $\approx 0.6\,{\rm keV}$ and $\approx 5\times 10^{-15}\,{\rm erg\,cm^{-2}\,s^{-1}}$ for BH process, respectively.  We can observe that the maximum flux generated by secondary pairs in the photo-pion process is $\sim$ three orders of magnitude larger than the BH process. We show in Figure \ref{sed} the contribution of synchrotron emission from the photo-pion process (dotted-dashed line).  The solid black line in this Figure represents the total contribution (SSC model, seed photons, and the synchrotron radiation from secondary pairs).   Given the distance from the blazar TXS 0506+056 to Earth, photons above 20 TeV are drastically suppressed due to EBL absorption. Taking into account the effect of this absorption described in \cite{2008A&A...487..837F},  the attenuation factor $\exp\{-\tau(E_\gamma, z) \}$ to the photon flux  lies in the range of $\sim$ 70 - 260.\\    
With the parameters reported in Table \ref{table2}, the proton luminosity associated to the ``neutrino flare"  is $\approx 3.5\times 10^{46}\,{\rm erg\,s^{-1}}$ which is slightly lower than the Eddington luminosity.  Therefore, from eq.~(\ref{Lp}), the proton density associated to this flare becomes $211\,{\rm cm^{-3}}$. This value is very similar to the electron density found with our model after fitting the archival data. We can conclude that within the dissipation region of the jet there is one cold proton per electron \citep[i.e., it has neutral charge][]{2011ApJ...736..131A, 2013ApJ...768...54B, 2017APh....89...14F}.\\
We conclude that during September 2014 and March 2015, pairs were continuously created outside the disk forming a sub-relativistic outflow with velocities in the range of $0.3\lesssim\beta_p\lesssim0.7$ as predicted in numerical simulations \citep{1999MNRAS.305..181B}.   During this period, the annihilation lines reached the dissipation region and interacted with the accelerating protons, producing the 10 - 20 TeV neutrinos detected in the IceCube telescope.\\
The expected MeV gamma-ray contributions from synchrotron  BH pair production and annihilation lines directly impact the broadband SEDs.  Nearby blazars including TXS 0506+056 would be potential candidates for MeV gamma-ray orbiting observatories such as  All-sky Energy Gamma-ray observatory \citep[AMEGO;][]{2019arXiv190707558M}  and enhanced ASTROGAM \citep[e-ASTROGAM; ][]{2018JHEAp..19....1D} which will explore the sky in the energy bands of 0.3 - 100 MeV and 0.3 - 3 GeV, respectively.\\
Our results indicate that for a typical Lorentz factor of $\sim 20$, the maximum energy of the characteristic neutrino is $\sim 20\,{\rm TeV}$.  For sources with a redshift of $z=0.5$ and a typical Lorentz factor $\Gamma_{\rm b} < 30$, we expect the characteristic neutrino energy in the range of 30 - 40 TeV. Therefore,  neutrino events with energies $E_{\nu}\lesssim 40\, {\rm TeV}$ reported  by the IceCube Collaboration \citep{2017arXiv171001191I, 2014PhRvL.113j1101A}  might be explained through this phenomenological model.\\
Neutrino multiplet as detected on February 17, 2016, by the IceCube \citep{2017A&A...607A.115I}  could be expected from this process and would be promising sources for IceCube-Gen2 \citep{2014arXiv1412.5106I}.\\

%
%
\section{Discussion and Summary}
We studied the high-energy neutrino production in the inner outflows of blazars.  As seed photons,  we considered the radiation from a sub-relativistic pair plasma scenario and a relativistic plasma in Wein equilibrium state. This radiation emitted from the photosphere can be detected above $> 511\,{\rm keV}$ up to a few MeV on Earth (observer frame), and only at dozens of keV in the frame of the dissipation region. These keV photons reach the dissipation region and interact with accelerating protons producing pion decay products ($\gamma$, $\nu_{\rm \mu,e}$ and $e^\pm$).  Depending on the parameter values of the dissipation region and the sub-relativistic and the relativistic pair plasma scenario,   gamma-rays, neutrinos and synchrotron photons generated by secondary pairs could be detected by orbiting and ground telescopes on Earth.\\
We want to emphasize that \cite{2014PhRvD..90b3007M} calculated the neutrino production considering seed photons originated in the blazar zone, the gas clouds, and the accretion disk.  The main novelty of this work is to consider the seed photons as those generated by the annihilation pairs released from the photosphere in different pair plasma scenarios.\\
We calculated the neutrino and gamma-ray spectra from extragalactic BL Lac objects with redshifts $z<2$. We showed that  the neutrino flux from the sub-relativistic pair plasma scenario has a significant contribution in the 1 - 30 TeV energy range, and also that synchrotron photons produced by the secondary pairs  could be observed in MeV gamma-ray orbiting satellites such as eASTROGAM and AMEGO.\\
As a particular case, we applied our model to describe the broadband SED of the blazar TXS 0506+056 successfully during the quiescent state.  We estimated the neutrino spectrum from the sub-relativistic pair plasma.  In this case, VHE neutrinos in the 10 - 20 TeV energy range can be detected without its electromagnetic counterpart, as observed during the ``neutrino flare" reported  between September 2014 and March 2015 by the IceCube collaboration. Based on previous models  \cite[i.e.][]{2019NatAs...3...88G},  we concluded that the ``neutrino flare" reported in  2014 - 2015 had a different origin to the flare in 2017 detected in all the electromagnetic bands.\\ 
Neutrino multiplet as detected on February 17, 2016, by the IceCube \citep{2017A&A...607A.115I}  could be expected from this process and would be promising sources for IceCube-Gen2 \citep{2014arXiv1412.5106I}. For sources with a redshift of $z=0.5$ and typical Lorentz factors $\Gamma_{\rm b} < 30$, we expect the characteristic neutrino energies in the range of 30 - 40 TeV. Therefore,  neutrino events with energies $E_{\nu}\lesssim 40\, {\rm TeV}$ reported  by the IceCube collaboration \citep{2017arXiv171001191I, 2014PhRvL.113j1101A}  might be explained through this phenomenological model.\\

\section*{Acknowledgements}

We  thanks  Antonio Marinelli,  Shan Gao and Kohta Murase for useful discussions.  This work was supported by UNAM-DGAPA-PAPIIT through  grant  IA102019

\section{DATA AVAILABILITY STATEMENT}

The data underlying this article will be shared on reasonable request to the corresponding author.







\newpage

\begin{table}
\centering \renewcommand{\arraystretch}{0.5}\addtolength{\tabcolsep}{1pt}
\caption{Values used to compute the Neutrino Flux and the synchrotron emission from secondary pairs}
\label{table1}
\begin{tabular}{l  c  c }
 \hline \hline
\scriptsize{} &  \scriptsize{Parameter}  &\hspace{0.5cm}   \scriptsize{Value}     \\ 

\hline \hline
\scriptsize{Blob region}\\\hline

\scriptsize{Bulk Lorentz factor}                               &  \scriptsize{$\Gamma_{\rm b}$}  &\hspace{0.5cm} \scriptsize{10}		\\	
\scriptsize{Variability (day)}    		 	        &  \scriptsize{$t$}  &\hspace{0.5cm} \scriptsize{1}\\	
\scriptsize{Proton luminosity ($L_{\rm edd}$)}      		 	        & \scriptsize{$L_{\rm p}$}  &\hspace{0.5cm} \scriptsize{0.5}\\	 
\scriptsize{Magnetic field (G)}      		 	        & \scriptsize{$B$ (G)}  &\hspace{0.5cm} \scriptsize{1}\\	 
\scriptsize{Maximum proton energy (PeV)}      		 	        & \scriptsize{$E_{\rm max}$ }  &\hspace{0.5cm} \scriptsize{1}\\	 
\scriptsize{Spectral index}      		 	        & \scriptsize{$\alpha_{\rm p}$}  &\hspace{0.5cm} \scriptsize{2.1}\\	 
			        
\\ \hline
A sub-relativistic pair plasma scenario \\ \hline
\scriptsize{Bulk Lorentz factor}                               &  \scriptsize{$\Gamma_{\rm p}$}  &\hspace{0.5cm} \scriptsize{1.4}		\\	
\scriptsize{Relative Lorentz factor}                               &  \scriptsize{$\Gamma_{\rm rel}$}  &\hspace{0.5cm} \scriptsize{4.2}		\\	
\scriptsize{Photosphere radius ($r_{\rm g}$)}                               &  \scriptsize{$r_{\rm ph}\, $}  &\hspace{0.5cm} \scriptsize{2}		\\	

\\
\hline 
A relativistic plasma in Wein equilibrium state\\ \hline
\scriptsize{Bulk Lorentz factor}                               &  \scriptsize{$\Gamma_{\rm p}$}  &\hspace{0.5cm} \scriptsize{1.8}		\\	
\scriptsize{Relative Lorentz factor}                               &  \scriptsize{$\Gamma_{\rm rel}$}  &\hspace{0.5cm} \scriptsize{3.1}		\\	
\scriptsize{Photosphere radius ($r_{\rm g}$)}                               &  \scriptsize{$r_{\rm ph}\,$}  &\hspace{0.5cm} \scriptsize{3}		\\	
\\
\hline \hline
\end{tabular}
\end{table}
%

%
\begin{table*}
\centering \renewcommand{\arraystretch}{0.8}\addtolength{\tabcolsep}{1pt}
\caption{Values found in estimating the best-fit curve  of the archival data points with our  model.}
\label{table2}
\begin{tabular}{l  c  c }
 \hline \hline
\scriptsize{} &  \scriptsize{Parameter}  &\hspace{0.8cm}   \scriptsize{Value}     \\ 

\hline \hline
Fitting parameters\\\hline

\scriptsize{Bulk Lorentz factor}                               &  \scriptsize{$\Gamma_{\rm p}$}  &\hspace{0.5cm} \scriptsize{$19\pm3$}		\\	
\scriptsize{Dissipation region (${\rm cm}$)}    		 	        &  \scriptsize{$r_d$}  &\hspace{0.5cm} \scriptsize{$(3.3\pm 0.41)\times 10^{16}$}\\	
\scriptsize{Electron density (${\rm cm^{-3}}$)}      		 	        & \scriptsize{$N_{\rm e}$}  &\hspace{0.5cm} \scriptsize{ $(2.2\pm0.5)\times 10^{2}$}\\	 
\scriptsize{Magnetic field (${\rm G}$)}      		 	        & \scriptsize{$B$}  &\hspace{0.5cm} \scriptsize{$0.16\pm0.03$}\\	 
\scriptsize{Low-energy electron spectral Index}       & \scriptsize{$\alpha_{\rm 1}$}  &\hspace{0.5cm} \scriptsize{$2.47\pm0.02$}\\
\scriptsize{Medium-energy electron spectral Index}       & \scriptsize{$\alpha_{\rm 2}$}  &\hspace{0.5cm} \scriptsize{$3.69\pm0.14$}\\
\scriptsize{High-energy electron spectral Index}       & \scriptsize{$\alpha_{\rm 3}$}  &\hspace{0.5cm} \scriptsize{$4.50\pm0.15$}\\
\\ \hline
Derived parameter\\ \hline
\scriptsize{Minimum electron Lorentz factor$^a$}                               &  \scriptsize{$\gamma_{\rm e, min}$}  &\hspace{0.5cm} \scriptsize{$800$}		\\	
\scriptsize{Break 1 electron Lorentz factor}                               &  \scriptsize{$\gamma_{\rm e, c1}$}  &\hspace{0.5cm} \scriptsize{$(6.26\pm0.71))\times10^3$}		\\	
\scriptsize{Break 2 electron Lorentz factor}                               &  \scriptsize{$\gamma_{\rm e, c2}$}  &\hspace{0.5cm} \scriptsize{$(4.75\pm 0.61)\times10^4$}		\\	
\scriptsize{Maximum electron Lorentz factor}                               &  \scriptsize{$\gamma_{\rm e, max}$}  &\hspace{0.5cm} \scriptsize{$(1.16\pm 0.12)\times10^8$}		\\	
\scriptsize{Magnetic Luminosity ($\rm erg\,s^{-1}$)}                               &  \scriptsize{$L_{\rm B}$}  &\hspace{0.5cm} \scriptsize{$(3.73\pm 0.59)\times10^{43}$}		\\	
\scriptsize{Electron Luminosity ($\rm erg\,s^{-1}$)}                               &  \scriptsize{$L_{\rm e}$}  &\hspace{0.5cm} \scriptsize{$(1.91\pm0.33)\times10^{46}$}		\\	

\hline \hline
$^a$ This value was not derive from our model, but rather used an input.
\end{tabular}
\end{table*}
%
%
%
\newpage


\bsp	
\label{lastpage}
\end{document}